\documentclass{pasj00}

\draft

\begin{document}

\SetRunningHead{Y. Takeda et al.}{Surface oxygen abundances of red giants}
\Received{2014/12/08}
\Accepted{2014/12/18}

\title{Can surface oxygen abundances of red giants\\
be explained by the canonical mixing theory?\thanks{
Based on data collected at Okayama Astrophysical Observatory
of NAOJ (Okayama, Japan).}
}

%

\author{
Yoichi \textsc{Takeda,}\altaffilmark{1,2}
Bun'ei \textsc{Sato,}\altaffilmark{3}
Masashi \textsc{Omiya,}\altaffilmark{1}
 and
Hiroki \textsc{Harakawa}\altaffilmark{1}
}

\altaffiltext{1}{National Astronomical Observatory, 2-21-1 Osawa, 
Mitaka, Tokyo 181-8588}
\email{takeda.yoichi@nao.ac.jp}
\altaffiltext{2}{The Graduate University for Advanced Studies, 
2-21-1 Osawa, Mitaka, Tokyo 181-8588}
\altaffiltext{3}{Tokyo Institute of Technology, 2-12-1 Ookayama, 
Meguro-ku, Tokyo 152-8550}

\KeyWords{stars: abundances --- stars: atmospheres ---  
stars: evolution --- stars: late-type 
}

\maketitle

\begin{abstract}
Extensive oxygen abundance determinations were carried out for
239 late-G/early-K giant stars of 1.5--5 $M_{\odot}$ by applying
the spectrum-fitting technique to O~{\sc i} 7771--5 and 
[O~{\sc i}] 6300/6363 lines based on the high-dispersion spectra
in the red region newly obtained at Okayama Astrophysical Observatory.
Our main purpose was to clarify whether any significantly large 
($\ltsim$~0.4--0.5~dex) O-deficit really exists in these evolved stars, 
which was once suspected by Takeda et al. (2008, PASJ, 60, 781) from 
the analysis of the [O~{\sc i}] 5577 line, since it (if real) is
inexlainable by the current theory and may require the necessity of 
special non-canonical deep mixing in the envelope. 
We found, however, that the previous [O/H]$_{5577}$ results 
(differential abundances relative to the sun) were 
systematically underestimated compared to the more reliable 
[O/H]$_{7773}$ (from O~{\sc i} 7771--5 triplet lines) or [O/H]$_{6300}$ 
(from [O~{\sc i} 6300 line) obtained in this study. 
Comparing the updated [O/Fe] ratios with the theoretically 
predicted surface abundance changes caused by mixing of nuclear-processed 
products dredged-up from the interior, we concluded that the oxygen 
deficiency in these red giants is insignificantly marginal (only by 
$\ltsim 0.1$~dex), which does not contradict the expectation 
from the recent theoretical simulation. This consequence of reasonable 
consistency between theory and observation also applies to the extent 
of peculiarity in [C/Fe] and [Na/Fe], which were also examined by 
reanalyzing the previous equivalent-width data of C~{\sc i} 5052/5380 
and Na~{\sc i} 6160 lines.
\end{abstract}

%


\section{Introduction}

Red giants are stars of low-to-intermediate mass, which have already evolved 
off the main sequence (after the exhaustion of core hydrogen) while increasing 
their radius. Since the surface temperature progressively drops down according to
this evolution, the deep convection zone is developed. As a result, some portion 
of the H-burning (CNO cycle) product processed in the interior may be salvaged 
and mixed into the envelope, by which the surface abundances of specific 
light elements are more or less altered. According to the current theory 
of envelope mixing, it is primarily the CN-cycled (C$\rightarrow$N reaction) 
material that is dredged up, while the product of ON-cycle (O$\rightarrow$N 
reaction; occurring in deeper region of higher $T$) is unlikely to cause 
any significant abundance changes since the mixing does not penetrate 
into such a deep/hot layer. Thus, the predicted surface abundances 
are characterized by a deficit in C as well as an enhancement in N
(with its peculiarity degree increasing with mass/luminosity), 
while O is practically unaffected (see, e.g., Fig. 24 in Mishenina et al. 2006).
Yet, although such expected tendency has been almost confirmed observationally 
for C and N, the situation for O seems to remain still more controversial.

For example, Mishenina et al. (2006) found that 
the oxygen abundances (determined based on the [O~{\sc i}] forbidden line at 
6300~$\rm\AA$) in red-clump giants of $\sim$~1--3~$M_{\odot}$ are almost normal 
($\langle$[O/Fe]$\rangle \simeq 0$), in agreement with the theoretical 
prediction. Further, Tautvai\u{s}ien\.{e} et al.'s (2010) similar study
on red-clump giants resulted in essentially the same conclusion. 
These results suggest the practical validity
of the current mixing theory, in the sense that it satisfactorily explains
the behaviors of all C, N, and O.

However, Takeda, Sato, and Murata (2008; hereinafter 
referred to as Paper~I) reported based their extensive 
spectroscopic analysis of 322 late-G and early-K giants 
(4500~K $\ltsim T_{\rm eff} \ltsim$ 5500~K, 
1.5 $\ltsim \log g \ltsim$ 3.5, 1 $\ltsim M/M_{\odot} \ltsim$ 5) 
that [O/Fe] (determined from [O~{\sc i}] 5577 line) showed a subsolar 
tendency with the extent of peculiarity increasing with $M$ (cf. figure 12 
therein). Moreover, this O-deficiency seemed to be correlated with a 
deficit of C as well as with an overabundance of Na. This may imply that 
a significant mass-dependent deep dredge-up of (not only CN-cycle but also) 
ON-cycle products may take place in the envelope of red giants. If this 
is real, it would possibly require a substantial revision of the current theory 
for the dredge-up in the envelope of red giants. 
  
One serious concern regarding the consequence in Paper~I was, however, that 
only one weak forbidden [O~{\sc i}] line of $\chi_{\rm low} = 1.97$~eV 
at 5577~$\rm\AA$ (which had been rarely used in previous studies)
had to be invoked for the O abundance determination, since only
spectra of $\sim$~5000--6200~$\rm\AA$ region were available at that time.
As a matter of fact, we noticed based on the comparison with the results
of other studies published so far that the oxygen abundances determined 
from this line do not necessarily agree with those derived from
the [O~{\sc i}] 6300 line (most popularly used for O-abundance analyses
of red giants) or O~{\sc i} 7771--5 triplet lines (which were also
shown to be a good indicator of oxygen abundances by Takeda et al. 1998), 
as described in the Appendix of Paper~I. Given this situation, we have 
realized the necessity of reinvestigating the oxygen abundances of red giants 
by using the presumably more reliable [O~{\sc i}] 6300 (along with similar
[O~{\sc i} 6363) and O~{\sc i} 7771--5 lines based on an extensive sample 
of stars like the case of Paper~I.

Motivated by this requirement, we decided to revisit the oxygen problem 
for red giants based on new observational material obtained for an extensive 
sample of 239 stars (making up $\sim$~3/4 of all the targets in Paper~I), 
which cover the longer wavelength region (up to $\sim 8800$~$\rm\AA$) 
and allowed us to use such O~{\sc i} and [O~{\sc i}] lines mentioned above. 
In addition, in order to revise the abundances of C and Na (important
key elements affected by dredge-up of nuclear-processed products) along with O, the equivalent widths 
of [O~{\sc i}] 5577, C~{\sc i} 5052/5380, Na~{\sc i} 6160 lines measured 
in Paper~I were also reanalyzed in a more careful manner (e.g., taking into account 
the CO formation effect or the non-LTE effect), and a new spectrum-fitting 
analysis was performed for the [C~{\sc i}] 8727 line by using our new spectra.
The purpose of this paper is to report the outcome of this analysis.

\section{Observational data and stellar parameters}

Our spectroscopic observations were done in 2012 August, 2013 May, 
2013 August, 2013 November, and 2013 December by using the HIDES 
(HIgh Dispersion Echelle Spectrograph) placed at the coud\'{e} focus 
of the 188 cm reflector at Okayama Astrophysical Observatory.
Equipped with three mosaicked 4K$\times$2K CCD detectors\footnote{
Note that, until 2006 when the data used in Paper~I were obtained, 
only one 4K$\times$2K CCD was available and the wavelength coverage 
was only $\sim$~1200~$\rm\AA$ in HIDES.}
at the camera focus, HIDES enabled us to obtain an echellogram covering 
5100--8800~$\rm\AA$ (wavelength span of $\sim 3700$~$\rm\AA$) with a 
resolving power of $R \sim 67000$ (case for the normal slit width 
of 200~$\mu$m) in the mode of red cross-disperser. 
While we tried to cover as many stars of the 322 objects studied 
in Paper~I as possible, we could finally get data for 239 stars 
(about $\sim$~3/4) over these five observing periods.

The reduction of the spectra (bias subtraction, flat-fielding, 
scattered-light subtraction, spectrum extraction, wavelength 
calibration, and continuum normalization) was performed by using 
the ``echelle'' package of the software IRAF\footnote{
IRAF is distributed by the National Optical Astronomy Observatories,
which is operated by the Association of Universities for Research
in Astronomy, Inc. under cooperative agreement with the National 
Science Foundation.} in a standard manner. 
Since $\sim$~2--3 consecutive frames (mostly 10--20 min exposure for each) 
were observed in a night for each star in many cases, we co-added 
these to improve the signal-to-noise ratio, by which the average S/N 
of most stars turned out to be in the range of $\sim$~100--300. 
Regarding the solar spectrum used as a reference, we adopted
the moon spectra observed with HIDES, which were published by 
Takeda et al. (2005).

Some representative spectra in four wavelength regions 
comprising our main target lines ([O~{\sc i}] 6300, [O~{\sc i}] 6363,\footnote{
Since this line overlaps the broad and shallow autoionization wing 
of the Ca~{\sc i} 6361.786 line, the normalization was done by treating
this autoionization wing as a pseudo-continuum.}
O~{\sc i} 7771--5, and [C~{\sc i}] 8727) are shown in figure 1
for four objects  (HD~4627, HD~19476, HD~71369, and the sun).

We remark that the line [O~{\sc i}] 6300 critically falls near to 
the inter-detector gap. Unfortunately, since our choice of the 
cross-disperser setting angle was slightly inappropriate due to 
our mis-adjustment in 2013 August and 2013 November observations, this 6300 
line fell just outside of the detector and could not be used for 
these data (this incident applied to 78 stars out of 239 in total).

Regarding the stellar parameters ($T_{\rm eff}$, $\log g$, [Fe/H], 
$v_{\rm t}$, $M$, $\ldots$) and the atmospheric model of each star,
we exclusively adopted those determined/used in Paper~I unchanged.
The mutual correlations of the parameters (including the theoretical 
HR diagram along with the evolutionary tracks) are plotted in figure 2. 
The basic data of our 239 program stars (HD number, stellar parameters, 
date of observation) are given in the electronic table~E (``tableE.txt''). 

\section{Abundance determination}

\subsection{Synthetic spectrum fitting}

Abundance determinations were carried out by using our spectrum-analysis 
tool MPFIT, which was developed by Y. Takeda based on Kurucz's (1993a) 
WIDTH9 code. This program establishes the most optimum solutions
accomplishing the best match between theoretical and observed spectra
by using the numerical algorithm described in Takeda (1995), 
while simultaneously varying the abundances of 
relevant key elements ($\log \epsilon_{1}$, $\log \epsilon_{2}$, $\ldots$), 
the macrobroadening parameter ($v_{\rm M}$),\footnote{
This $v_{\rm M}$ is the $e$-folding half-width of 
the Gaussian broadening function ($\propto \exp[-(v/v_{\rm M})^{2}]$),
which represents the combined@effects of instrumental broadening, 
macroturbulence, and rotational velocity
(cf. subsubsection 4.2.2 in Paper I).} 
and the radial-velocity (wavelength) shift ($\Delta \lambda$).

Specifically, our spectrum fitting was conducted for the following four 
wavelength regions, 7770--7777~$\rm\AA$  
[to determine $\log\epsilon$(O) from the O~{\sc i} 7771--5 lines
and $\log\phi$(CN) from the CN lines (see the note in table 1 for the meaning
of $\log\phi$(CN))], 6300--6302~$\rm\AA$ 
[to determine $\log\epsilon$(O) from the [O~{\sc i}] 6300 line
(partially blended with the Ni~{\sc i} 6300 line)], 
6362--6365~$\rm\AA$ [to determine $\log\epsilon$(O) from the
[O~{\sc i}] 6363 line (partially blended with the CN 6363 line)], 
and 8726--8730~$\rm\AA$ [to determine $\log\epsilon$(C) from 
the [C~{\sc i}] 8727 line partially blended with the Fe~{\sc i} 8727 line],
as summarized in table 1. Note that the procedure of analysis for 
the 7770--7777~$\rm\AA$ region was done in essentially the same manner 
as in Takeda et al. (1998), which should be consulted for more details.
Since the O~{\sc i} 7771--5 permitted lines are known to suffer 
a considerable non-LTE correction (in contrast to other [O~{\sc i}] or 
[C{\sc i}] forbidden lines, for which LTE is guaranteed to hold), we 
explicitly took the non-LTE effect into consideration in the calculation 
of these triplet lines following Takeda (2003).   

The atomic parameters (wavelengths, excitation potentials, oscillator 
strengths) of important spectral lines adopted in this fitting are 
presented in table 2. As for the damping parameters (which are unimportant
in the present case because very strong lines are absent in the
relevant wavelength regions), the data given in Kurucz and Bell (1995) 
were used; if not available therein, we invoked the default 
treatment of Kurucz's (1993a) WIDTH9 program. 

The convergence of the solutions turned out fairly successful  
for most of the cases. How the theoretical spectrum for the converged 
solutions fits well with the observed spectrum for each star is 
displayed in figure 3 (7770--7777~$\rm\AA$ fitting), 
figure 4 (6300-6302~$\rm\AA$ fitting and 6362-6365~$\rm\AA$ fitting), 
and figure 5 (8726--8730~$\rm\AA$ fitting).

\subsection{Equivalent widths and abundance uncertainties}

While the synthetic spectrum fitting directly yielded the abundance 
solutions of O (and C), this approach is not necessarily suitable 
when one wants to quantify the contribution of blending components, 
evaluate the extent of the non-LTE correction, or to study the 
abundance sensitivity to changing the atmospheric parameters (i.e., 
it is rather tedious to repeat the fitting process again and 
again for different assumptions or different atmospheric parameters).
Therefore, with the help of Kurucz's (1993a) WIDTH9 program\footnote{
This WIDTH9 program had been considerably modified in various 
respects; e.g., inclusion of non-LTE effects, treatment of total 
equivalent width for multi-component lines; etc.}, 
we computed the equivalent widths corresponding to the relevant lines
 ``inversely'' from the abundance solutions (resulting from spectrum 
synthesis) along with the adopted atmospheric models and parameters, 
which are much easier to handle:
$W_{{\rm O}7771}$, $W_{{\rm O}7774}$, and $W_{{\rm O}7775}$ 
(for O~{\sc i}~7771, 7774, 7775) from $\log\epsilon$(O) of 7771--7777~$\rm\AA$ fitting;
$W_{{\rm O}6300}$ and $W_{{\rm Ni}6300}$ (for [O~{\sc i}]~6300 and Ni~{\sc i}~6300)
from $\log\epsilon$(O) and $\log\epsilon$(Ni) of 6300--6302~$\rm\AA$ fitting;
$W_{{\rm O}6363}$ and $W_{{\rm CN}6363}$ (for [O~{\sc i}]~6363 and CN~6363)
from $\log\epsilon$(O) and $\log\phi$(CN) of 6362--6365~$\rm\AA$ fitting; 
$W_{{\rm C}8727}$ and $W_{{\rm Fe}8727}$ (for [C~{\sc i}]~8727 and Fe~{\sc i}~8727)
from $\log\epsilon$(C) and $\log\epsilon$(Fe) of 8726--8730~$\rm\AA$ fitting.
Regarding the O~{\sc i} 7771, 7774, 7775 lines, the non-LTE as well as LTE abundances 
were also derived based on such evaluated $W$ values, from which the non-LTE corrections 
($\Delta^{\rm NLTE}_{7771}$, $\Delta^{\rm NLTE}_{7774}$, $\Delta^{\rm NLTE}_{7775}$)
were computed. 
The results of the abundances,\footnote{
Abundances are given in $\log\epsilon$(X) as well as in [X/H], where $\log\epsilon$(X) 
is the logarithmic abundance of element X in the usual normalization of 
$\log\epsilon$(H) = 12, and [X/H] is the differential abundances relative to the Sun
defined as [X/H] $\equiv \log \epsilon_{*}({\rm X}) - \log \epsilon_{\odot}({\rm X})$.
} the equivalent widths ($W$), and the non-LTE corrections 
($\Delta^{\rm NLTE}$) are summarized in the on-line table~E (``tableE.txt''). 
Besides, such derived $W$ values (along with the related quantities) for the relevant 
lines are plotted against $T_{\rm eff}$ and [Fe/H] in figure 6, from which we can 
realize the relative importance of the contribution of blended lines
as compared to the O and C lines of our interest.

Regarding the abundance errors due to ambiguities in atmospheric parameters,
we estimated the changes in $\log\epsilon$(O) and $\log\epsilon$(C)
by repeating the analysis on the $W$ value of each line 
($W_{{\rm O}7774}$, $W_{{\rm O}6300}$, $W_{{\rm O}6363}$, $W_{{\rm C}8727}$)
while perturbing the standard atmospheric parameters interchangeably by $\pm 100$~K 
in $T_{\rm eff}$, $\pm 0.2$~dex in $\log g$, and $\pm 0.2$~km~s$^{-1}$ 
in $\xi$ (which are considered to be typical magnitudes of ambiguities; 
see subsection 3.1 in Paper~I, especially the comparison with the literature
values shown in figures 5--7 therein). The resulting abundance changes are 
summarized in table 3, from which the following tendencies are read.\\
--- The $T_{\rm eff}$-sensitivity of O~{\sc i} 7771--5 lines is appreciably large 
($\ltsim 0.2$~dex for a change of 100~K), while that of forbidden lines is 
insignificant (much smaller for [C~{\sc i}] 8727 and negligible for 
[O~{\sc i} 6300/6363).\\  
--- Regarding the effect of changing $\log g$, all these lines show
almost the same behaviors of mild sensitivity ($\sim 0.1$~dex
for a change of 0.2~dex).\\
--- Abundances are practically unaffected by a change in $v_{\rm t}$
($\sim$~0.02~dex for O~{\sc i} 7771--5 or a few hundredths dex for
forbidden lines\footnote{Regarding the [C~{\sc i}] 8727 forbidden line,  
the sign of the abundance change in response to varying $v_{\rm t}$
is contrary to what is intuitively expected (i.e., the resulting abundance 
slightly increases for a larger $v_{\rm t}$). This phenomenon is sometimes seen
in case of very weak lines of light elements, which may be interpreted 
as due to the difference of photon-forming layers at different points
of line profiles (see subsection 3.2 in Takeda 1994).}
in response to a variation of 0.2~km~s$^{-1}$).

\subsection{Reanalysis of C, O, and Na equivalent widths in Paper I}

While C, O, and Na abundances were determined for 322 giants from the equivalent 
widths of C~{\sc i}~5052/5380, [O~{\sc i}]~5577, and Na~{\sc i}~6160 lines 
in Paper I, the treatment adopted there was not necessarily 
full-fledged. More precisely, no consideration was made to the non-LTE effect 
as well as the molecule-formation effect, they are not quantitatively significant
for these lines and tend to be more or less cancelled in [X/H] (differential 
abundances relative to the Sun).
However, attention should be paid also to these effects (which generally differ 
from star to star) to accomplish as accurate [X/H] values as possible.
Accordingly, we decided to reanalyze the equivalent widths ($W$) for these lines 
(published in Paper I) while taking these factors explicitly into account 
(as we have done in our main analysis described in subsection 3.1), in order 
to revise [C/H]$_{5052/5380}$, [O/H]$_{5577}$, and [Na/H]$_{6160}$ for 
our 239 program stars. 

Computing populations by including molecules (where the formation of CO is 
especially important) was done by following the treatment used in Kurucz's 
(1993) ATLAS9 program. Meanwhile, the non-LTE corrections were evaluated 
as done in Takeda and Honda (2005) for C and Takeda et al. (2003) for Na 
(while LTE is essentially valid for the [O~{\sc i}] 5577 forbidden line). 
The resulting non-LTE corrections are plotted against $W$ in figures 7a (C~{\sc i} 5052), 
7b (C~{\sc i} 5380), and 7c (Na~{\sc i} 6160), where the tendency of increasing 
$|\Delta^{\rm NLTE}|$ with an increase in $W$ is seen. Since the extents of
(negative) non-LTE corrections for red giants are generally larger than that 
for the sun (reflecting the nature of lower-density atmospheres), 
[C/H] as well as [Na/H] have been corrected in the downward direction
as shown figures 7d and 7f, though the corrections ($\ltsim 0.1$~dex) are 
quantitatively not very important. On the other hand, since $\log\epsilon$(O)
derived from [O~{\sc i}]~5577 line tends to be somewhat increased by including
CO formation effect (which is more appreciable for lower-$T_{\rm eff}$ stars),
[O/H]$_{5577}$ values have been revised in the upward direction for stars with
$T_{\rm eff} \ltsim 5000$~K (cf. figure 7e), though these corrections
($\ltsim 0.06$~dex) are again not so significant. These updated results 
for [C/H]$_{5052/5380}$ (which is the average of [C/H]$_{5052}$ and 
[C/H]$_{5380}$), [O/H]$_{5577}$, and [Na/H]$_{6160}$ are
also summarized in electronic table E (tableE.txt). 

\section{Discussion}

\subsection{Comparison of the abundances from different lines}

We have thus derived the oxygen abundances for the program stars 
by applying the spectrum-fitting technique to three line features: 
O~{\sc i} 7771--5, [O~{\sc i}] 6300, and [O~{\sc i}] 6363 lines
(subsection 3.1). How are they compared with each other, and 
how do they relate to the results from O~{\sc i} 5577 (subsection 3.3) 
derived by reanalyzing the $W$ values of Paper I?
Which abundance indicator is most reliable?

The mutual comparisons of these [O/H] values 
are depicted in figures 8a--8f, in which we can recognize that a tendency 
[O/H]$_{5577} \ltsim$ [O/H]$_{6363} <$ [O/H]$_{6300} \simeq$ [O/H]$_{7773}$
roughly holds.  That is, a reasonable consistency is observed between 
[O/H]$_{7773}$ and [O/H]$_{6300}$ (cf. figure 8d), while [O/H]$_{6363}$
and (especially) [O/H]$_{5577}$ are systematically lower than these two.

We consider that the O~{\sc i} 7771--5 feature is the most reliable 
abundance indicator among these in the present case of red giants, given that 
the spectrum-synthesis technique is applied and the non-LTE effect 
(rather appreciable; cf. figure 6b) is properly taken into account, 
because it comprises three components of different strengths and 
the blending effect by other lines is insignificant. 
That is, we may regard that the resulting abundance is reliable 
if the whole triplet feature could be adequately fitted by theoretically synthesized 
spectrum, even though its large $T_{\rm eff}$-sensitivity (cf. subsection 3.2) 
may be a drawback.

Regarding the forbidden lines [O~{\sc i}] 6300 and [O~{\sc i}] 6363, they are
often used for O-abundance determination of red giants, since they get 
strengthened in lower-$g$ atmospheres and formed essentially in LTE. 
We consider, however, that their credibility would be comparatively lower,
because only one [O~{\sc i}] line component is blended by line(s) of 
other species and the removal of this effect inevitably causes loss of accuracy.  
The contamination is especially significant for the [O~{\sc i}] 6363.78 line
which is overlapped by the CN 6363.78 line. Actually, the contribution of 
this CN line can be appreciable (typically several tens per cent or 
even more up to a comparable level; cf. figure 6e and figure 6h).
Moderately enhanced CN population by up to $\ltsim 0.4$~dex (see figure 6c) 
may also be partly responsible for this fact, which is a combined result of 
considerably increased N and mildly decreased C in the atmosphere of 
red giants (see, e.g., Mishenina et al. 2006). 
Meanwhile, concerning the [O~{\sc i}] 6300.30, we can see from figure 6d and 
figure 6g that the contribution of the blending Ni~{\sc i} 6300.35 line is 
not very significant compared to the case of [O~{\sc i}] 6363.

Given these results, it is reasonable to state that O~{\sc i} 7771--5 and 
[O~{\sc i}] 6300 lines are the reliable abundance indicators in the present
case, and thus we should adopt the O abundances derived from these two features
(which are mostly in agreement as shown in figure 8d).
 
\subsection{Problem involved with the [O I] 5577 line}

This judgement naturally leads to a conclusion that [O/H]$_{5577}$ values are 
erroneously inadequate, since they are systematically smaller than [O/H]$_{7773}$ 
(figure 8e) and [O/H]$_{6300}$ (figure 8a). Actually, this possibility was already 
suspected in Paper~I (cf. Appendix therein); but at that time we could not find 
a reason why this [O~{\sc i}] 5577 line yielded wrong differential oxygen 
abundances relative to the sun.

After Paper~I has been published, however, we noticed Mel\'{e}ndez and 
Asplund's (2008) work, who studied the solar oxygen abundance based on 
the [O~{\sc i}] 5577 line by using a 3D model atmosphere.
We realized from their paper that this [O~{\sc i}] feature is contaminated
by P$_{1}$27 and P$_{1}$26 lines of C$_{2}$ (1--2) Swan band (cf. their Fig. 2), 
the blending effect of which needs to be taken into account.

We tried to estimate how much contribution is made by these C$_{2}$ lines
to the strength of the [O~{\sc i}] 5577 feature. Since these two lines are 
almost the same strength (P$_{1}$27 is at 5577.338~$\rm\AA$
with $\chi_{\rm low}$ = 0.726~eV and $\log gf = 1.364$, P$_{1}$26 is at 
5577.404~$\rm\AA$ with $\chi_{\rm low}$ = 0.727~eV and $\log gf = 1.381$)
according to Kurucz's (1993b) molecular line data (c2da.dat), we may tentatively 
focus only on the former 5577.338 line. Regarding the [O~{\sc i}] line, 
the same atomic data as used in Paper I  was adopted (5577.339~$\rm\AA$, 
$\chi_{\rm low}$ = 1.967~eV and $\log gf = -8.204$). 
We computed the equivalent widths of these lines ([O~{\sc i} 5577.339 and 
C$_{2}$ 5577.338) for the sun and a typical red giant ($T_{\rm eff} = 4900$~K, 
$\log g = 2.5$, and [Fe/H] = 0) on the assumption of [C/H] = 0, and 
obtained (1.69~m$\rm\AA$ and 0.75~m$\rm\AA$) for the sun and 
(7.40~m$\rm\AA$ and 2.95~m$\rm\AA$) for a red giant.

It might appear that the relative contribution of C$_{2}$ blending is almost 
the same for the sun and red giant and cancelled in the net differential abundances,
since the resulting $W$(O):$W$(C$_{2}$) ratios are practically identical as
1.7:1.5($=2\times 0.75$) and 7.4:5.9($=2\times 2.95$),
However, we should keep in mind that C is generally deficient in the atmosphere
of red giants typically by 0.2--0.3~dex (cf. figure 11a and figure 11b)
by mixing of CN-cycled products. As the population of C$_{2}$ molecules
scales as $n({\rm C}_{2}) \propto \epsilon ({\rm C})^{2}$, the contribution
of C$_{2}$ lines to the 5577 feature must be further reduced by a factor of
$\sim$~3--4 to an insignificant level compared with the [O~{\sc i}] line itself. 

Thus, our interpretation for the reason why we obtained erroneously 
low [O/H]$_{5577}$ values is simply that we did not take into account the 
blending effect of two C$_{2}$ lines in our analysis; its influence was
not so significant for red giants but serious for the sun (standard star).
That is, since the reference solar O abundance was overestimated by $\sim 0.3$~dex 
by neglecting this effect in the analysis of the [O~{\sc i}] (+ C$_{2}$) line 
feature at 5577~$\rm\AA$ (while the extent of such an overestimation is much 
milder or even negligible for giant stars), we eventually obtained systematically 
underestimated [O/H]$_{5577}$ in comparison to (presumably correct) [O/H]$_{7773}$ 
or [O/H]$_{6300}$.
 
\subsection{Correlation between [O/Fe], [C/Fe], and [Na/Fe]}

Let us examine based on the revised abundances how O, C, and Na are correlated 
with each other, and how the situation is changed compared to what was concluded 
in Paper I. We adopt the oxygen abundances derived from O~{\sc i} 7771--5 and 
[O~{\sc i}] 6300 lines according to the discussion in subsection 4.1,
while the non-LTE abundances obtained by reanalyzing the 
$W$(Na~{\sc i}~6161) data are used for Na (subsection 3.2). 

Regarding C, although the abundances from the [C~{\sc i}] 8727 forbidden line
were newly determined in this study, they suffer from appreciable contamination 
of the overlapping Fe~{\sc i} line (cf. figure 6f and figure 6i), and considered 
to be less reliable. Since the consistency between the [C/H] values derived from 
[C~{\sc i}] 8727 and those from C~{\sc i}~5052/5380 is not necessarily satisfactory 
(i.e., rather large scatter with some systematic difference; cf. figure 8g), 
we put larger weight to the latter non-LTE abundances obtained by reanalyzing 
the $W$(C~{\sc i}~5052) and $W$(C~{\sc i}~5380) data (subsection 3.2), 
where the results derived from these two permitted C~{\sc i} lines were 
averaged since they are in agreement with each other (figure 8h).

The mutual correlations between [O/Fe], [C/Fe], and [Na/Fe], and their dependence 
upon [Fe/H] as well as $M$ are depicted in figure 9.
Comparing figure 9a ([O/Fe] vs. [C/Fe]), figure 9b ([Na/Fe] vs. [C/Fe]),
and figure 9c ([Na/Fe] vs. [O/Fe]) with the corresponding figures in Paper I
(figures 12a, 12b, and 12c therein), we can see that only [O/Fe] values
have been appreciably shifted upward by $\sim 0.3$~dex on the average 
(due to the use of O~{\sc i} 7771--5 and [O~{\sc i}] 6300 lines in this study 
instead of the [O~{\sc i}] 5577 line in our previous work),
while no significant changes are seen in [C/Fe] as well as in [Na/Fe]. 
That is, oxygen does not show remarkable deficiency any more, as seen from 
the range of $-0.2 \ltsim$~[O/Fe]~$\ltsim +0.5$ (in contrast to the previous 
$-0.5 \ltsim$~[O/Fe]~$\ltsim +0.2$). 

The reason why we suspected in Paper I the existence of non-canonical 
mixing was that the O-deficiency appeared to conform to the C-deficiency
as well as to the Na-enrichment, and the extent of this peculiarity 
seemed to increase with the stellar mass (cf. figure 12 therein). 
We thus consider at that time that these characteristics may indicate 
a deep dredge-up of H-burning (ON-, CN-, NeNa-cycle) products. 

However, now that the revised results for O, C, and Na are established, 
we can reasonably interpret the abundance trends of these elements as follows.\\
--- The positive correlation between [O/Fe] and [C/Fe] is simply due to the 
$\alpha$-element-like behavior (i.e., [X/Fe] decreases with an increase
in [Fe/H]) shown by these elements (cf. figure 9d and figure 9e), which results
from the galactic chemical evolution as exhibited by nearby solar-type stars 
(see, e.g., Takeda \& Honda 2005).\\
--- The inversed correlation between Na and O (i.e., [Na/Fe] tends to increase 
with a decrease in [O/Fe]; figure 9c) is mostly attributed to the ``upturn'' 
nature of [Na/Fe] with an increase of [Fe/H], which contrasts the trend 
of O. That is, the different behavior of [O/Fe] (figure 9d) and [Na/Fe] (figure 9f) 
against a change of [Fe/H] is the cause of this anti-correlation.\\ 
--- The apparent variation of [O/Fe] against the stellar mass reported in Paper I 
([O/Fe] $\sim 0$ at $M\sim$~1--2~$M_{\odot}$ and [O/Fe] $\sim -0.4$ at 
$M\sim$~3--4~$M_{\odot}$) has been updated to [O/Fe] $\sim$~0.4--0.5 at 
$M\sim$~1--2~$M_{\odot}$ and [O/Fe] $\sim 0$ at $M\sim$~3--4~$M_{\odot}$ 
by the upward revision of [O/Fe], which thus can not be regarded 
as a $M$-dependence of O-deficiency any more. This trend is simply due to 
the fact that low-$M$ stars tend to have low [Fe/H] (cf. figure 2f), 
where [O/Fe] is generally enhanced up to $\sim$~+0.4--0.5 (figure 9d).

Accordingly, since we have significantly revised the [O/Fe] ratios for the program 
stars in this study, which are systematically higher (typically by $\sim$~0.3--0.4~dex)
than the values reported in Paper I, we came to a conclusion that the oxygen abundances
in red giants atmospheres do not show appreciable peculiarities and that the 
observed relations between [O/Fe], [C/Fe], and [Na/Fe] can be reasonably explained 
mostly by their intrinsic qualitative characteristics caused by galactic chemical 
evolution, without invoking any special a-posteriori mechanism (e.g., non-canonical 
deep mixing) such that generating a significant O-abundance anomaly. 

\subsection{Does theoretical predictions explain the observed abundance trends?}

Now, we return to the subject which motivated this investigation:
``Are the surface oxygen abundances of red giants consistent with the prediction 
from the canonical theory of envelope mixing? How about carbon and sodium?'' 
In order to answer this question, we have to carefully evaluate the changes 
in the surface abundances of these elements, which were caused by a mixing of 
nuclear-processed materials dredged-up in the red-giant stage. 
For this purpose, it is necessary to adequately take into account the chemical 
evolution effect (i.e., intrinsic [X/Fe] ratio of the gas at the time of star 
formation), which can be done by comparing the abundances of evolved red giants 
with those of unevolved dwarfs at the same metallicity.

In this discussion, we confine ourselves to the abundance results of red giants 
derived from permitted lines, which we consider to be most reliable: 
C abundances from C~{\sc i} 5052/5380 lines, O abundances from 
O~{\sc i} 7771--5 lines, and Na abundances from Na~{\sc i} 6160 line.
In addition, we also refer to the C, O, and Na abundances of FGK dwarfs 
(or subgiants) derived from the same lines, which were taken from
Takeda and Honda (2005) (for C and O) as well as Takeda (2007) (for Na).

The [X/Fe] vs. [Fe/H] diagrams (X = C, O, and Na) plotted for giants and dwarfs 
are shown in the left-side panels (a, b, and c) of figure 10, where the mean 
$\langle$[X/Fe]$\rangle$ at each metallicity group (0.1~dex bin within 
$-0.4 \le$~[Fe/H]~$\le +0.2$) along with the distribution of
$\langle$[X/Fe]$\rangle_{\rm giants} - \langle$[X/Fe]$\rangle_{\rm dwarfs}$ 
are also presented in the corresponding right-side panels (d, e, and f).
We can state from figure 10d, 10e, and 10f that the abundance changes
(compared to the initial values when stars were formed) suffered
in the red-giant phase by evolution-induced envelope mixing are
a moderate decrease of C by $\sim 0.2$~dex, only a slight decrease of O
by $\ltsim 0.1$~dex, and a marginal increase of Na by $\sim$~0.1--0.2 dex.

Then, what about the theoretically predicted abundance anomalies of
low-to-intermediate mass stars in the red giant phase? In figure 11 are shown 
the expected surface abundance changes of C, O, and Na during 
the course of post-main-sequence stellar evolution (plotted against $T_{\rm eff}$) 
calculated by Lagarde et al. (2012) for 1.5, 2.5, and 4 $M_{\odot}$ stars,
where the results for different assumptions of envelope mixing (standard 
treatment and treatment including rotational and thermohaline mixing)\footnote{
We consider that rotational/thermohaline mixing actually takes place 
in addition to the standard mixing of first dredge-up, according to Takeda 
and Tajitsu's (2014) recent study on the Be abundances of red giants.} 
are presented for two metallicity cases (0.3$\times$ solar metallicity and 
1$\times$ solar metallicity). 
Although it is difficult to confront these results in detail with the observations 
as the computed anomalies intricately depend on various factors (mass, metallicity, 
assumptions on mixing), we can draw the following consequences from the comparison 
of figure 10 and figure 11:\\
--- Regarding oxygen, our observational result (only a slight deficiency by 
$\ltsim 0.1$~dex) satisfactorily matches the theoretical expectation that 
the surface O abundances are hardly altered (the predicted decrease is 
$\ltsim 0.05$~dex at most). This means that the current theory for the mixing 
in the envelope of evolved stars is quite sufficient to account for
the observed oxygen abundances of red giants, without any necessity to invoke 
a non-canonical deep mixing causing a significant dredge-up of ON-cycle product.
\\
--- The observed mild enrichment of Na and deficiency of C are reasonably predicted 
by the simulations. That is, if we consider the typical case of a 2.5~$M_{\odot}$ 
star of solar-metallicity around $T_{\rm eff} \sim$~4800--5000~K, 
figure 11a$'$ and figure 11c$'$ suggest that the expected abundance changes
are an underabundance of C by $\sim$~0.2--0.3~dex and an overabundance of Na by
$\sim$~0.2--0.3~dex. Though the observed extents of anomaly ($\sim$~0.2~dex
deficiency for C and $\sim$~0.1--0.2~dex enrichment for Na) appear somewhat
smaller than the theoretical predictions (e.g., the case of Na), we may state
that theory and observation are tolerably consistent with each other. 

Consequently, according to what has been described above, the abundance 
characteristics of C (mildly deficient), O (barely changed or only slightly 
deficient), and Na (mildly enriched) observed in red giants are reasonably explained 
by the recent theoretical simulation such as that by Lagarde et al. (2012).
This consistency indicates that a substantial or intrinsic modification of the theory 
(such as an inclusion of special non-canonical deep mixing) is not necessary,
though some refinements on technical details in simulations (e.g., which kind 
of physical processes are to be included in the envelope mixing) will naturally 
be further in order.
     
\section{Summary and conclusion}

Takeda et al. (2008) suggested in Paper I based the analysis of the 
[O~{\sc i}] 5577 line carried out for 322 late G--early K giants that 
oxygen is significantly underabundant in their atmospheres.
If this is real, it might suggest a dredge-up of ON-cycle product 
caused by a non-canonical deep-mixing, such as that unable to be 
covered by the current theory.
 
However, this result apparently contradicted the consequence of other 
studies (e.g., Mishenina et al. 2006, Tautvai\u{s}ien\.{e} et al. 2010), 
which concluded based on the [O~{\sc i}] 6300 line that O is almost normal 
(without any sign of significant anomaly) in the atmosphere of red giants
in agreement with the theoretical prediction.  

In order to settle this issue by clarifying which conclusion represents the truth, 
extensive abundance determinations were conducted for oxygen (along with carbon 
and sodium) for 239 late-G/early-K giant stars by using various lines.
We applied the spectrum-fitting technique to O~{\sc i} 7771--5, [O~{\sc i}] 6300/6363, 
and [C~{\sc i}] 8727 lines to the red-region spectra newly obtained at 
Okayama Astrophysical Observatory, and reanalyzed the previously published 
equivalent widths of [O~{\sc i}] 5577, C~{\sc i} 5052/5380 and Na~{\sc i} 6160 lines.

It then revealed that the previous [O/H]$_{5577}$ results in Paper~I were 
systematically underestimated compared to the more reliable [O/H]$_{7773}$ 
(from O~{\sc i} 7771--5 triplet lines) or [O/H]$_{6300}$ (from [O~{\sc i}] 6300 
line) newly obtained in this study. Regarding the reason why
[O~{\sc i}] 5577 line yielded erroneously low [O/H], we consider 
that this was due to our neglect of the blending effect of C$_{2}$
lines in deriving the reference solar O abundance from this line.  

According to our updated results, the oxygen deficiency of these 
red giants is actually very marginal (only by $\ltsim 0.1$~dex), which is 
in good agreement with the expectation from the recent theoretical 
simulation by Lagarde et al. (2012).
The same conclusion also applies to the observed extents in the abundance
anomalies of C ($\sim$~0.2~dex deficit) as well as Na ($\sim$~0.1--0.2~dex 
enrichment) derived from C~{\sc i} 5052/5380 and Na~{\sc i} 6160 lines.

To sum up, the current theoretical simulations are considered to be 
successful enough in predicting the surface abundance changes of red giants, 
without any need of substantial revision (such as by incorporating 
an unreasonably deep mixing process) as far as O, C, and Na are concerned.

Finally, the fact that the atmospheric abundance of oxygen suffers little change 
even in the evolved giant stage would have a significant impact in observational
studies of galactic chemical evolution, since it means that intrinsically bright 
red giants may be exploited as a tracer of [O/Fe] ratio of the galactic gas 
at the time of star formation. 

\bigskip

Data reduction was in part carried out by using the common-use data analysis 
computer system at the Astronomy Data Center (ADC) of the National Astronomical 
Observatory of Japan.

\newpage
\setcounter{table}{0}
\begin{table}[h]
\caption{Treatment of relevant elemental abundances in the fitting.}
\begin{center}
\footnotesize
\begin{tabular}{cccl}\hline\hline
Abundance & LTE/NLTE & var/fix & Remark \\
\hline
\multicolumn{4}{c}{(7770--7777~$\rm\AA$ fitting)}\\
$\log\epsilon$(O) & non-LTE & varied & \\
$\log\epsilon$(Fe) & LTE & varied & \\
$\log\epsilon$(Nd) & LTE & varied & \\
$\log\phi$(CN) & LTE & varied & \\
\hline
\multicolumn{4}{c}{(6300--6302~$\rm\AA$ fitting)}\\
$\log\epsilon$(O) & LTE & varied & \\
$\log\epsilon$(Sc) & LTE & varied & \\
$\log\epsilon$(Fe) & LTE & varied & \\
$\log\epsilon$(Ni) & LTE & {\it fixed} & Taken from Paper I\\
\hline
\multicolumn{4}{c}{(6362--6365~$\rm\AA$ fitting)}\\
$\log\epsilon$(O) & LTE & varied & \\
$\log\epsilon$(Fe) & LTE & varied & \\
$\log\epsilon$(Zn) & LTE & varied & \\
$\log\phi$(CN)  & LTE & {\it fixed} &  Taken from 7770--7777~$\rm\AA$ fitting)\\
\hline
\multicolumn{4}{c}{(8726--8730~$\rm\AA$ fitting)}\\
$\log\epsilon$(C) & LTE & varied & \\
$\log\epsilon$(Si) & LTE & varied & \\
$\log\epsilon$(Fe) & LTE & varied & \\
\hline
\end{tabular}
\end{center}
\footnotesize
Note. 
The quantity $\phi$(CN) introduced in the fitting of 7770--7777~$\rm\AA$ 
and 6362--6365~$\rm\AA$ regions is a depth-independent factor, 
by which the occupation numbers of CN molecules (computed from a model 
atmosphere with metallicity-scaled CNO abundances) are to be multiplied 
to reproduce the observed  CN line strengths (cf. subsection 3.3 in Takeda et al. 1998). 
\end{table}

\setcounter{table}{1}
\begin{table}[h]
\caption{Atomic data of important spectral lines.}
\begin{center}
\footnotesize
\begin{tabular}{ccccl}\hline\hline
Species & $\lambda_{\rm air}$ & $\chi_{\rm low}$ & $\log gf$ & Source \\
        & ($\rm\AA$) & (eV) & (dex) &  \\
\hline
\multicolumn{5}{c}{(7770--7777~$\rm\AA$ fitting)}\\
O~{\sc i} & 7771.944 & 9.146& +0.32 & KB95 \\
O~{\sc i} & 7774.166 & 9.146& +0.17 & KB95 \\
O~{\sc i} & 7775.388 & 9.146& $-0.05$ & KB95 \\
Fe~{\sc i} & 7770.279 & 2.559 & $-5.12$ & TKS98 \\
Fe~{\sc i} & 7771.427 & 5.105 & $-2.48$ & TKS98 \\
Fe~{\sc i} & 7772.597 & 5.067 & $-2.16$ & TKS98 \\
Fe~{\sc i} & 7774.001 & 5.012 & $-2.27$ & TKS98 \\
Nd~{\sc ii}& 7773.052 & 0.000 & $-3.38$ & KB95 \\
CN         & 7770.76  & 1.18   & $-2.01$  & ET79 \\
CN         & 7772.88  & 1.28   & $-1.72$  & ET79 \\
CN         & 7772.95  & 1.31   & $-2.19$  & ET79 \\
CN         & 7775.42  & 1.16   & $-2.10$  & ET79 \\
CN         & 7776.68  & 1.36   & $-1.64$  & ET79 \\
CN         & 7776.69  & 1.13   & $-2.45$  & ET79 \\
\hline
\multicolumn{5}{c}{(6300--6302~$\rm\AA$ fitting)}\\
$[$O~{\sc i}$]$ & 6300.304 & 0.000 & $-9.72$ & GA97 \\
Ni~{\sc i}  & 6300.336 & 4.266 & $-2.43$ & TH05 \\
Sc~{\sc ii} & 6300.67  & 1.507 & $-1.84$ &  KB95 (wavelength adjusted) \\
Fe~{\sc i}  & 6301.498 & 3.654 & $-0.75$ & KB95 \\
\hline
\multicolumn{5}{c}{(6362--6365~$\rm\AA$ fitting)}\\
Zn~{\sc i}  & 6362.338 & 5.796 &  +0.15  & KB95 \\
Fe~{\sc i}  & 6362.885 & 4.186 & $-1.70$ & This study ($gf$ adjusted) \\
$[$O~{\sc i}$]$ & 6363.776 & 0.020 & $-10.19$ & GA97 \\
CN          & 6363.776 & 1.390 & $-1.75$  & KZ93 \\
Fe~{\sc i}  & 6364.360 & 4.795 & $-1.10$ & This study ($gf$ adjusted) \\
Fe~{\sc i}  & 6364.701 & 4.584 & $-1.92$ & This study ($gf$ adjusted) \\
\hline
\multicolumn{5}{c}{(8726--8730~$\rm\AA$ fitting)}\\
$[$C~{\sc i}$]$ & 8727.126 & 1.264 & $-8.21$ & KB95 \\
Fe~{\sc i}  & 8727.132 & 4.186 & $-3.93$ & KB95 \\
Si~{\sc i}  & 8728.010 & 6.181 & $-0.61$ & KB95 \\
Si~{\sc i}  & 8728.594 & 6.181 & $-1.72$ & KB95 \\
Fe~{\sc i}  & 8729.148 & 3.415 & $-2.95$ & KB95 \\
\hline
\end{tabular}
\end{center}
\footnotesize
Abbreviation code for the source of $gf$ values:
KB95 --- Kurucz and Bell (1995), TKS98 --- Takeda et al. (1998),
ET79 --- Eriksson and Toft (1979), GA97 --- Galav\'{\i}s et al. (1997),
TH05 --- Takeda and Honda (2005), KZ93 --- Kurucz (1993b). 
\end{table}

\setcounter{table}{2}
\begin{table}[h]
\caption{Abundance variations in response to changing atmospheric parameters.}
\begin{center}
\footnotesize
\begin{tabular}{ccccccc}\hline\hline
Line                  & $\Delta_{T+}$    & $\Delta_{T-}$    & $\Delta_{g+}$    & $\Delta_{g-}$    & $\Delta_{v+}$     & $\Delta_{v-}$    \\
\hline
 O~{\sc i} 7774       & $-0.166$ (0.012) & $+0.182$ (0.015) & $+0.090$ (0.006) & $-0.091$ (0.005) & $-0.021$ (0.008)  & $+0.020$ (0.008) \\
 $[$O~{\sc i}$]$ 6300 & $-0.005$ (0.006) & $+0.006$ (0.004) & $+0.096$ (0.002) & $-0.097$ (0.002) & $-0.004$ (0.001)  & $+0.003$ (0.002) \\
 $[$O~{\sc i}$]$ 6363 & $-0.005$ (0.007) & $+0.006$ (0.006) & $+0.096$ (0.003) & $-0.096$ (0.003) & $-0.001$ (0.001)  & $+0.001$ (0.001) \\
 $[$C~{\sc i}$]$ 8727 & $-0.067$ (0.020) & $+0.081$ (0.021) & $+0.116$ (0.008) & $-0.114$ (0.008) & $+0.003$ (0.002)  & $-0.006$ (0.003) \\
\hline
\end{tabular}
\end{center}
\footnotesize
Note. Changes of the abundances (expressed in dex) derived from each line 
in response to varying $T_{\rm eff}$ by $\pm 100$~K, $\log g$ by $\pm 0.2$~dex,
and $v_{\rm t}$ by $\pm 0.2$~km~s$^{-1}$. Shown are the mean values averaged over
each of the 239 stars, while those in parentheses are the standard deviations.
\end{table}

\newpage

\setcounter{figure}{0}
\begin{figure}
  \begin{center}
    \FigureFile(120mm,100mm){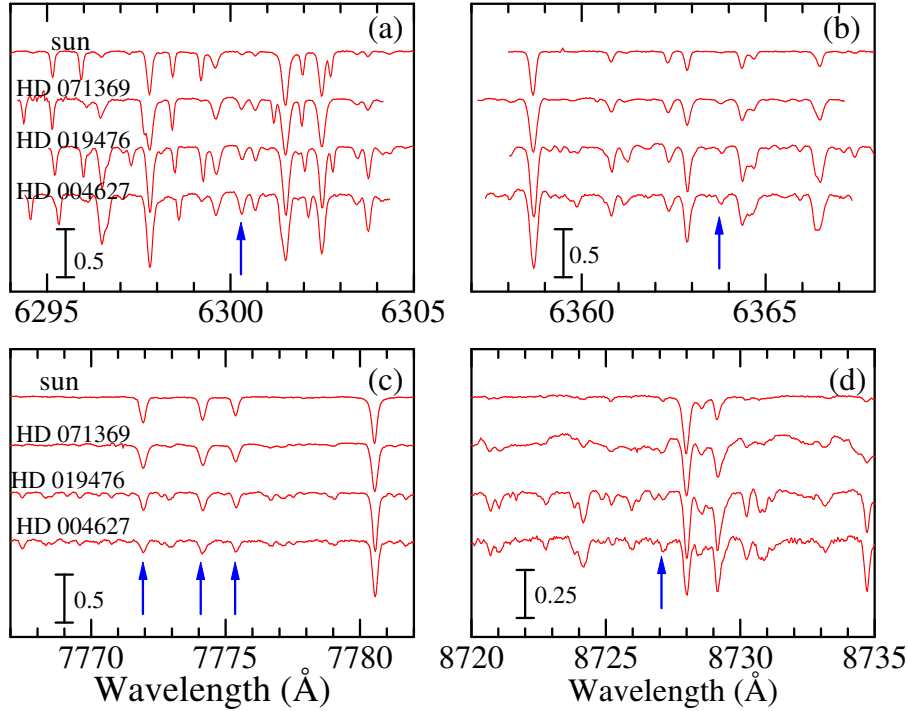}
  \end{center}
\caption{
Representative spectra of three stars selected from our sample 
(HD~71369, HD~19476, and HD~4627 with different $T_{\rm eff}$ of
5242~K, 4933~K, and 4599~K, respectively) and the sun (moon)
for the four wavelength regions comprising the lines used 
for our abundance determinations:
(a) $\cdots$ [O~{\sc i}]~6300, (b) $\cdots$ [O~{\sc i}]~6363, 
(c) $\cdots$ O~{\sc i}~7771--5, and (d) $\cdots$ [C~{\sc i}]~8727;
which are indicated by arrows in each panel.
Note that the ordinate scale of panel (d) is twice as magnified
as that of the other panels. 
}
\end{figure}

\setcounter{figure}{1}
\begin{figure}
  \begin{center}
    \FigureFile(120mm,160mm){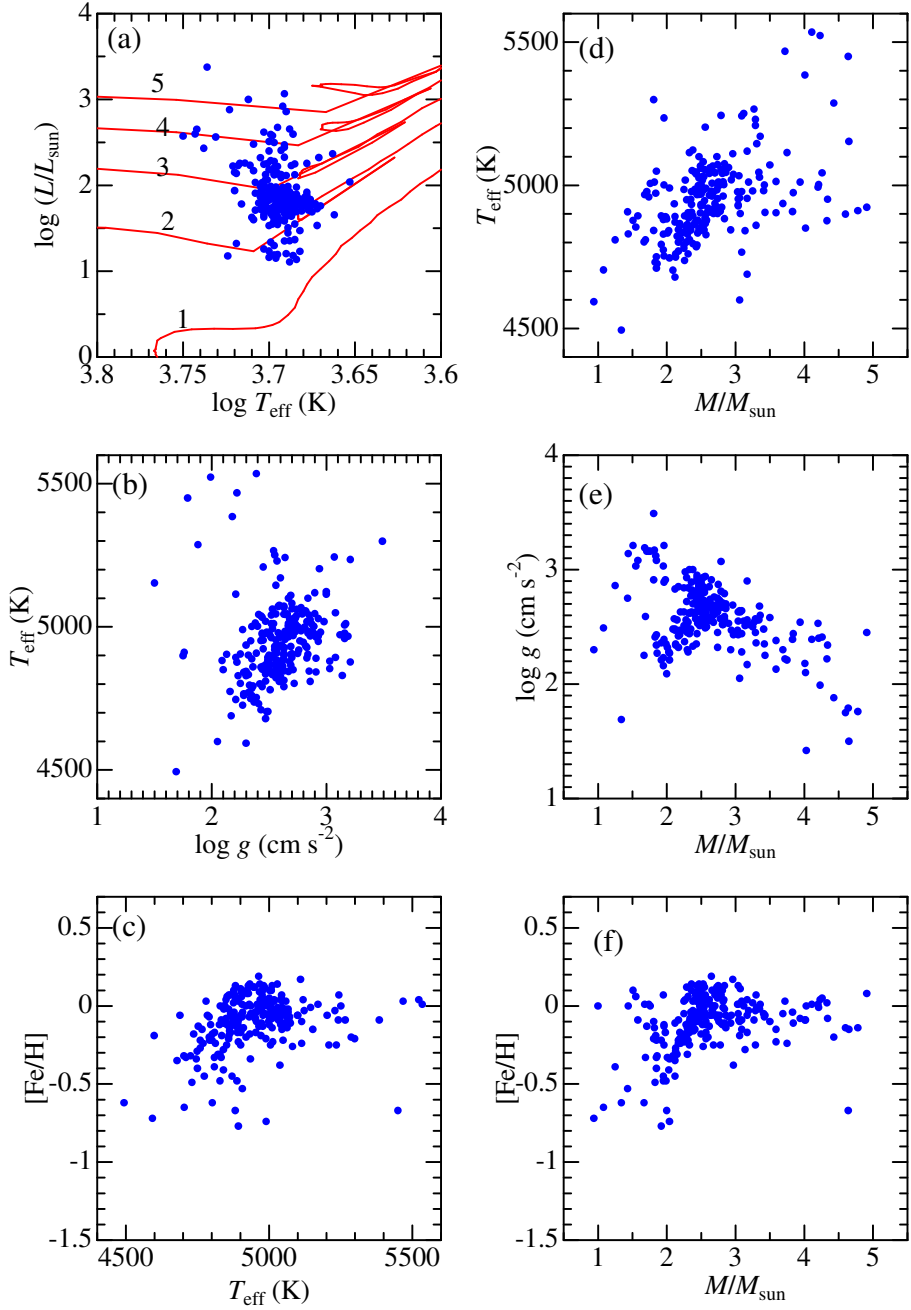}
  \end{center}
\caption{
Mutual correlations between the parameters of 239 program stars.
(a) $\log L$ vs. $\log T_{\rm eff}$, (b) $T_{\rm eff}$ vs. $\log g$,
(c) [Fe/H] vs. $T_{\rm eff}$, (d) $T_{\rm eff}$ vs. $M$, 
(e) $\log g$ vs. $M$, and (f) [Fe/H] vs. $M$.
In panel (a), Lejeune and Schaerer's (2001) theoretical evolutionary tracks of 
solar-metallicity stars (with masses of 1, 2, 3, 4, and 5 $M_{\odot}$) are 
also overplotted. 
}
\end{figure}

\setcounter{figure}{2}
\begin{figure}
  \begin{center}
    \FigureFile(160mm,200mm){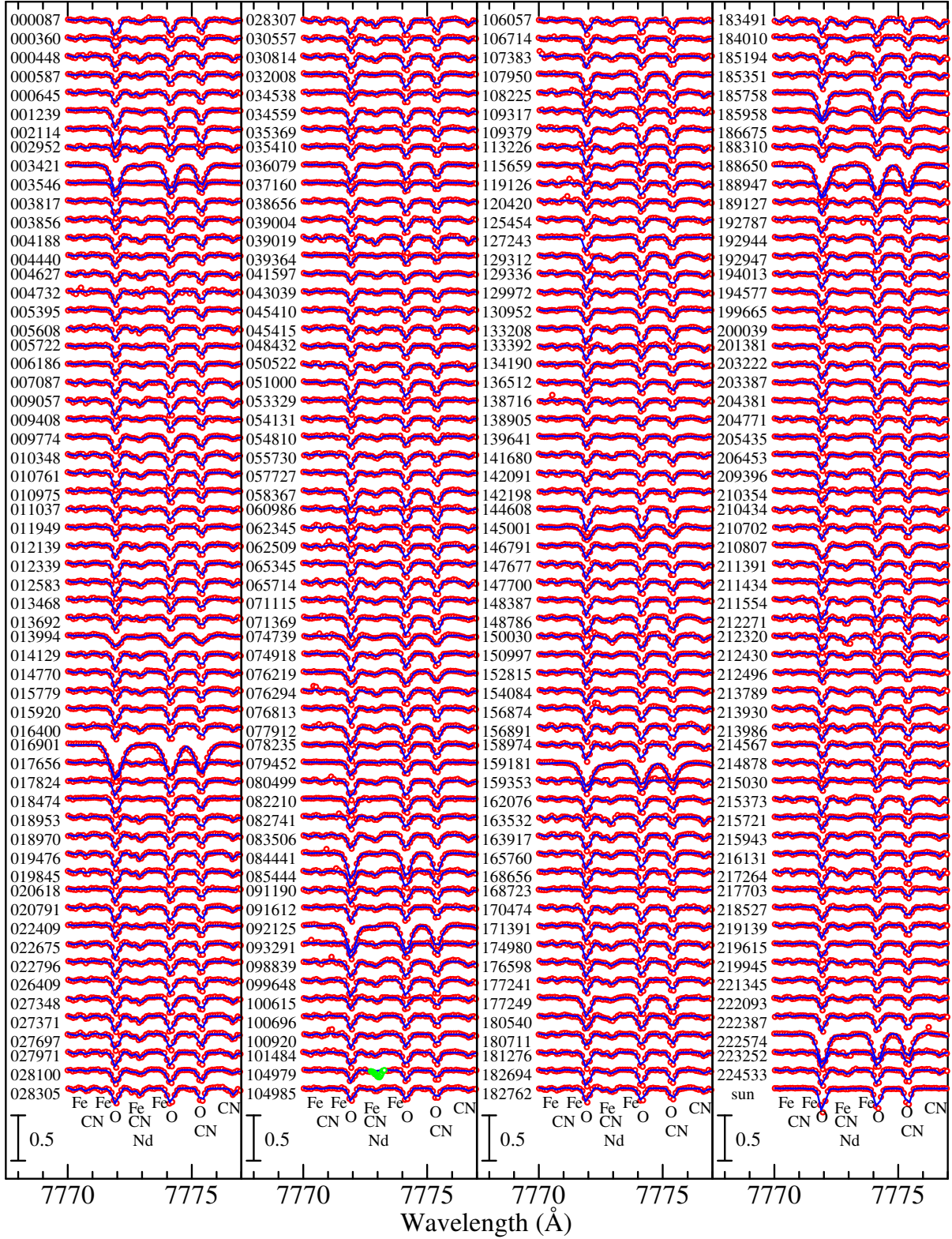}
  \end{center}
\caption{
Synthetic spectrum fitting in the 7770--7777~$\rm\AA$ region comprising 
the O~{\sc i} 7771--5 triplet lines. The best-fit theoretical spectra
are shown by blue solid lines, and the observed data are plotted
by red open circles (while those masked/disregarded in the fitting 
are highlighted in green). A vertical offset of 0.2 (in terms of the 
normalized flux with respect to the continuum) is applied to each 
relative to the adjacent ones. Each of the spectra are arranged 
in the increasing order of HD number (indicated on the left to each 
spectrum). The wavelength scale of each spectrum is adjusted to 
the laboratory system.
}
\end{figure}

\setcounter{figure}{3}
\begin{figure}
  \begin{center}
    \FigureFile(160mm,200mm){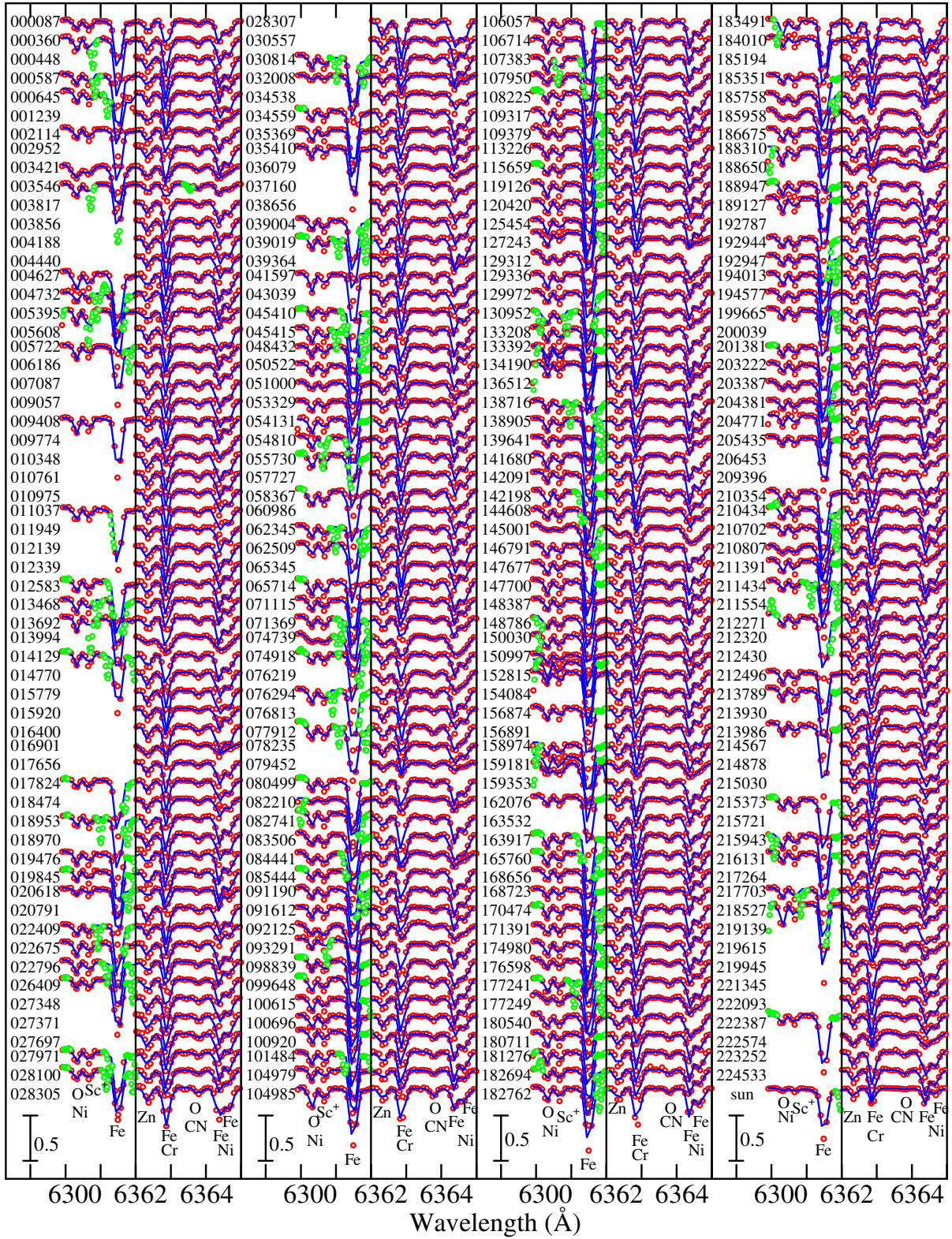}
  \end{center}
\caption{
Synthetic spectrum fitting in the 6300--6302~$\rm\AA$ region comprising 
the [O~{\sc i}] (+ Ni{\sc i}) 6300 line as well as 
in the 6362--6365~$\rm\AA$ region comprising 
the [O~{\sc i}] (+ CN) 6363 line. Otherwise, the same as in figure~3.
}
\end{figure}

\setcounter{figure}{4}
\begin{figure}
  \begin{center}
    \FigureFile(160mm,200mm){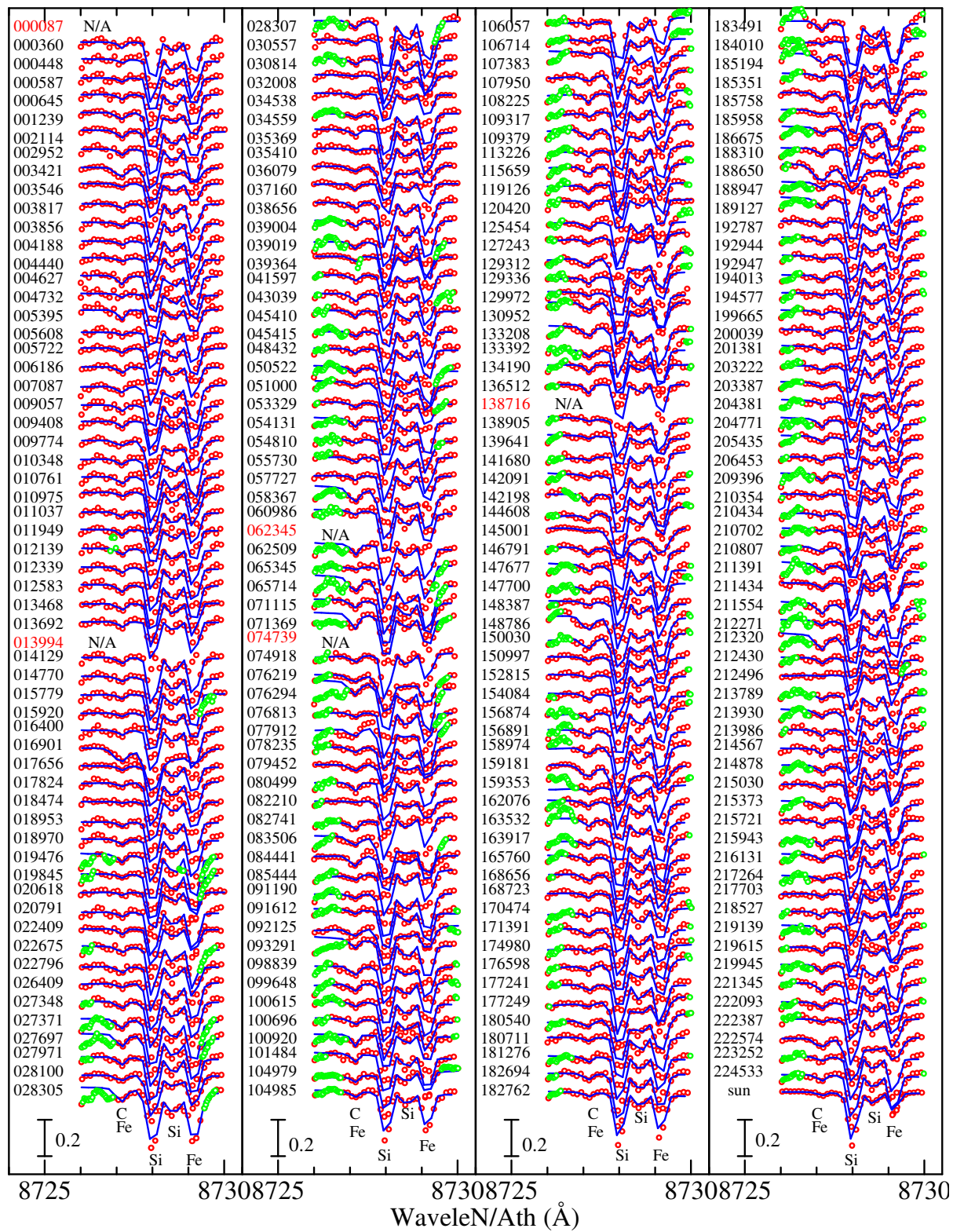}
  \end{center}
\caption{
Synthetic spectrum fitting in the 8726--8730~$\rm\AA$ region comprising 
the [C~{\sc i}] (+ Fe~{\sc i}) 8727 line. Otherwise, the same as in figure 3,
except that the applied vertical offset for each spectrum is 0.1 
(instead of 0.2).
}
\end{figure}

\setcounter{figure}{5}
\begin{figure}
  \begin{center}
    \FigureFile(160mm,200mm){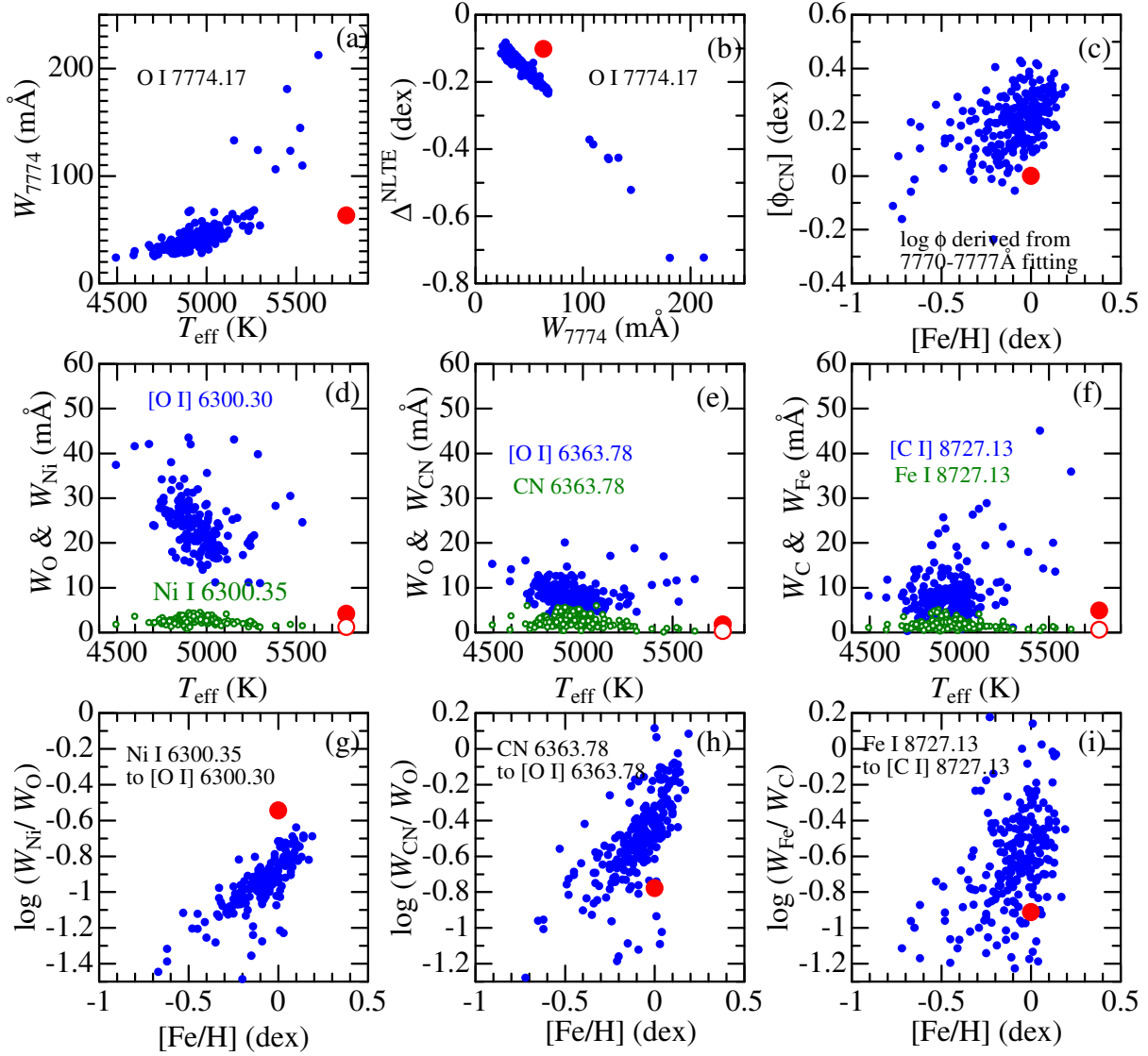}
  \end{center}
\caption{
Characteristics and behaviors of the quantities related to our abundance determinations.  
(a) $W_{7774}$ vs. $T_{\rm eff}$, (b) $\Delta_{7774}^{\rm NLTE}$ vs. $W_{7774}$,
(c) [$\phi_{\rm CN}$] ($\equiv \log \phi_{*} - \log \phi_{\odot}$) vs. [Fe/H],
(d) $W_{6300}$([O~{\sc i}] or Ni~{\sc i}) vs. $T_{\rm eff}$,
(e) $W_{6363}$([O~{\sc i}] or CN) vs. $T_{\rm eff}$,
(f) $W_{8727}$([C~{\sc i}] or Fe~{\sc i}) vs. $T_{\rm eff}$,
(g) $\log (W_{{\rm Ni}6300}/W_{{\rm O}6300})$ vs. [Fe/H],
(h) $\log (W_{{\rm CN}6363}/W_{{\rm O}6363})$ vs. [Fe/H], and
(i) $\log (W_{{\rm Fe}8727}/W_{{\rm C8727}})$ vs. [Fe/H].
The solar values are indicated by large red circles.
}
\end{figure}

\setcounter{figure}{6}
\begin{figure}
  \begin{center}
    \FigureFile(120mm,160mm){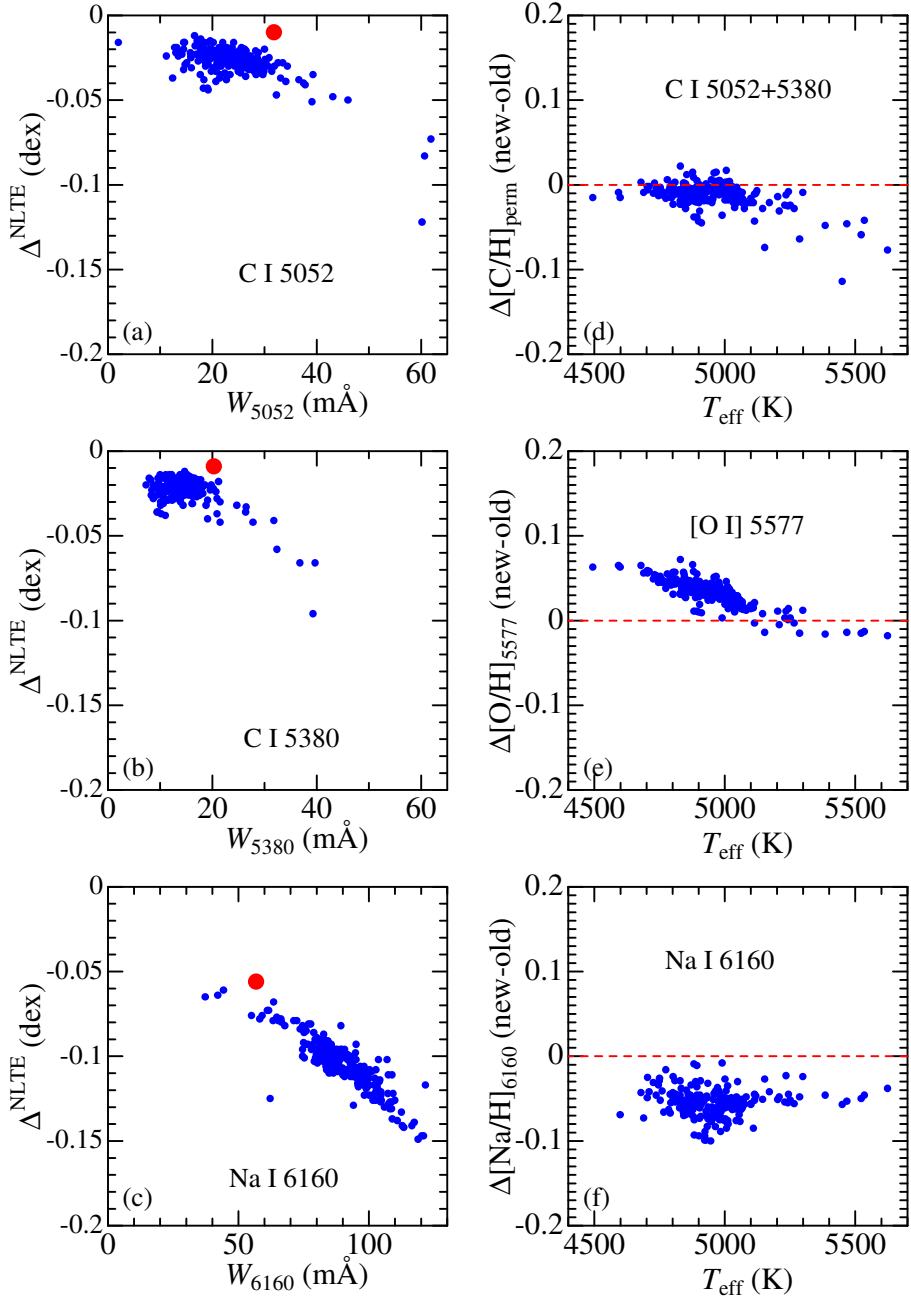}
  \end{center}
\caption{
Quantities related to our reanalysis of the equivalent widths measured in Paper~I.
Left panels (a, b, c): $W_{\lambda}$-dependence the non-LTE corrections
for C~{\sc i}~5052, C~{\sc i}~5380, and Na~{\sc i}~6160 lines,
where the results for the sun are represented by large red circles.
Right panels (d, e, f): difference of the new [C/H]$_{5052/5380}$, 
[O/H]$_{5577}$, and [Na/H]$_{6160}$ (relative to the previous values
published in Paper~I) plotted against $T_{\rm eff}$. 
}
\end{figure}

\setcounter{figure}{7}
\begin{figure}
  \begin{center}
    \FigureFile(110mm,150mm){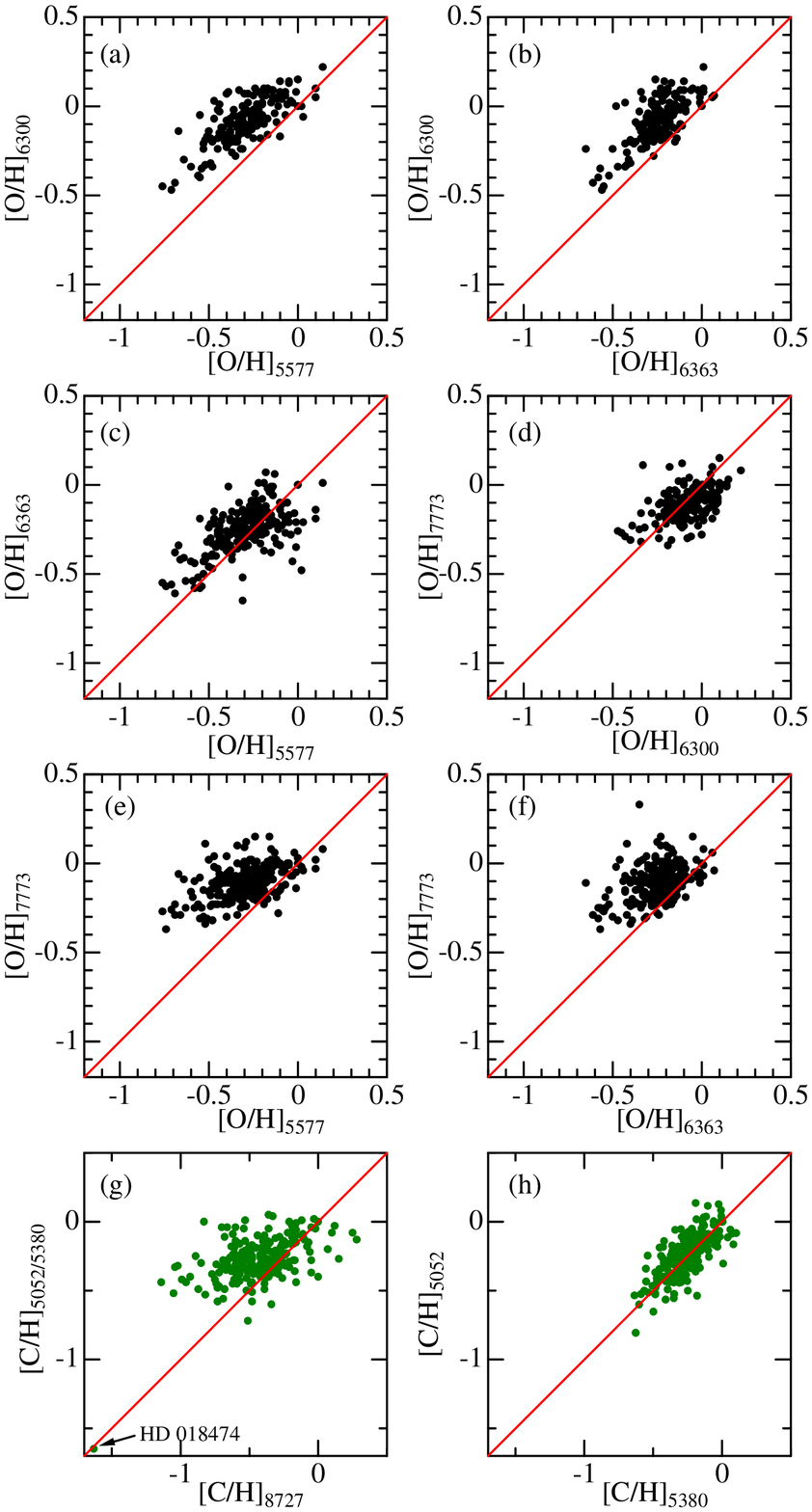}
  \end{center}
\caption{
Comparison of the various [O/H] results (as well as [C/H]) derived from different lines. 
(a) [O/H]$_{6300}$ vs. [O/H]$_{5577}$, (b)  [O/H]$_{6300}$ vs. [O/H]$_{6363}$,
(c) [O/H]$_{6363}$ vs. [O/H]$_{5577}$, (d)  [O/H]$_{7773}$ vs. [O/H]$_{6300}$,
(e) [O/H]$_{7773}$ vs. [O/H]$_{5577}$, (f)  [O/H]$_{7773}$ vs. [O/H]$_{6363}$,
(g) [C/H]$_{5052/5380}$ vs. [C/H]$_{8727}$, and (h)  [C/H]$_{5052}$ vs. [C/H]$_{5380}$.
}
\end{figure}

\setcounter{figure}{8}
\begin{figure}
  \begin{center}
    \FigureFile(160mm,200mm){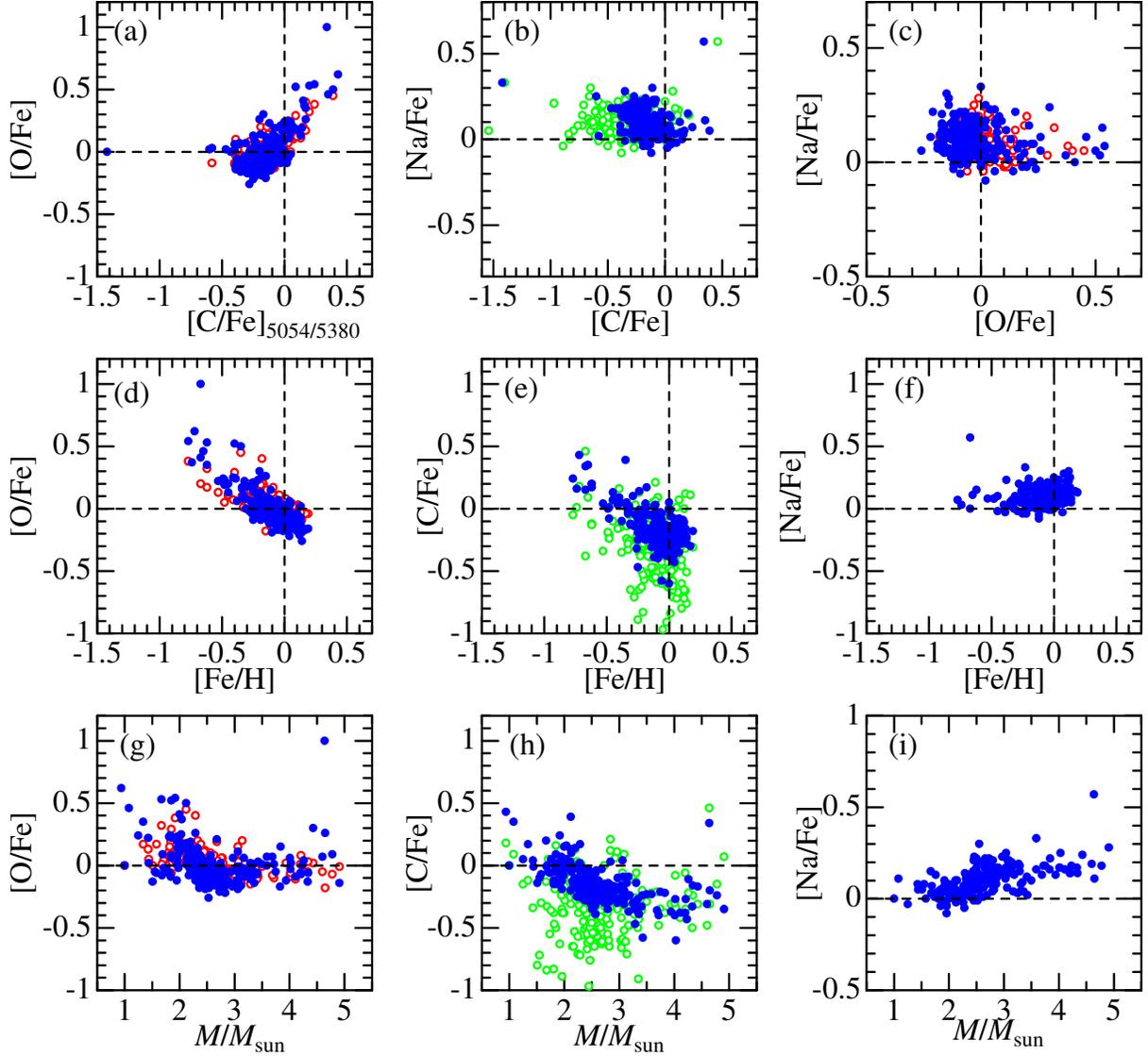}
  \end{center}
\caption{
Comparisons between [O/Fe] (from O~{\sc i} 7771--5 or [O~{\sc i} 6300), 
[C/Fe] (from C~{\sc i} 5052/5380 or [C~{\sc i}] 8727), and 
[Na/Fe] (from Na~{\sc i} 6160) with each other, and their relations
to [Fe/H] as well as $M$. 
(a) [O/Fe] (7771--5 or 6300) vs. [C/Fe] (5052/5380),  
(b) [Na/Fe] vs. [C/Fe] (5052/5380 or 8727),
(c) [Na/Fe] vs. [O/Fe] (7771--5 or 6300),
(d) [O/Fe] (7771--5 or 6300) vs. [Fe/H],
(e) [C/Fe] (5052/5380 or 8727) vs. [Fe/H],
(f) [Na/Fe] vs. [Fe/H],
(g) [O/Fe] (7771--5 or 6300) vs. $M$,
(h) [C/Fe] (5052/5380 or 8727) vs. $M$, and 
(j) [Na/Fe] vs. $M$.
The results from forbidden lines are shown in
open symbols in panels (a), (b), (c), (d), (e), (g), and (h), 
while those from permitted lines are in filled symbols.
}
\end{figure}

\setcounter{figure}{9}
\begin{figure}
  \begin{center}
    \FigureFile(120mm,160mm){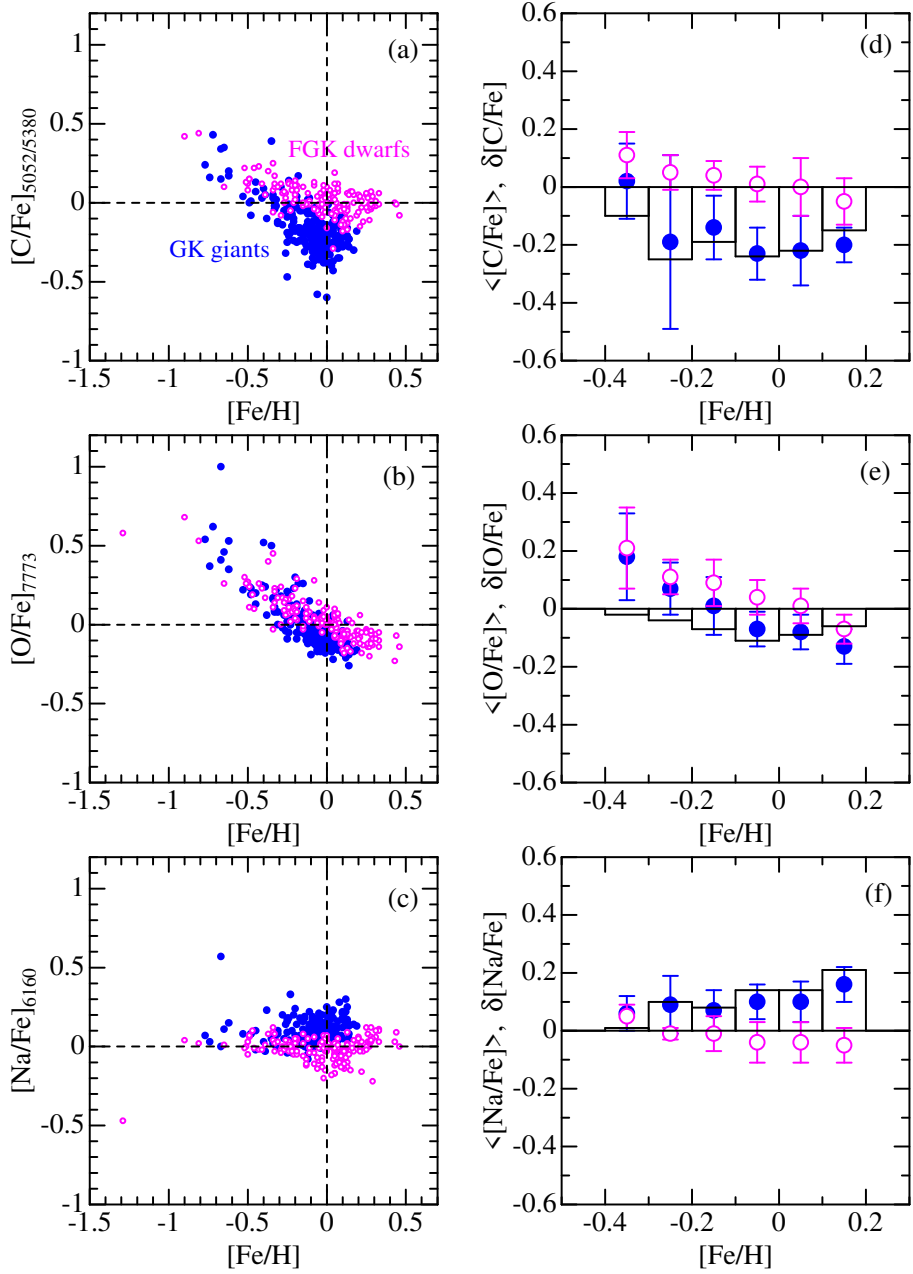}
  \end{center}
\caption{
Left-hand panels show the comparison of [X/Fe] vs.[Fe/H] relations 
(X = C or O or Na; based on permitted lines of C~{\sc i} 5052/5380, 
O~{\sc i} 7771--5, and Na~{\sc i} 6160) derived in this study for 
239 G--K giants (filled symbols) with those of 160 FGK dwarfs (open symbols)
published by Takeda and Honda (2005) (for C and O) and Takeda (2007) (for Na).
Symbols in the right-side panels give the mean $\langle$[X/Fe]$\rangle$ at 
each metallicity group (0.1~dex bin within $-0.4 \le$~[Fe/H]~$\le +0.2$) 
where error bars denote the standard deviations, while bar graphs represent 
the mean abundance differences between giants and dwarfs defined as 
$\langle$[X/Fe]$\rangle_{\rm giants} - \langle$[X/Fe]$\rangle_{\rm dwarfs}$. 
Panels (a)/(d), (b)/(e), and (c)/(f) correspond to C, O, and Na, respectively.
}
\end{figure}

\setcounter{figure}{10}
\begin{figure}
  \begin{center}
    \FigureFile(120mm,160mm){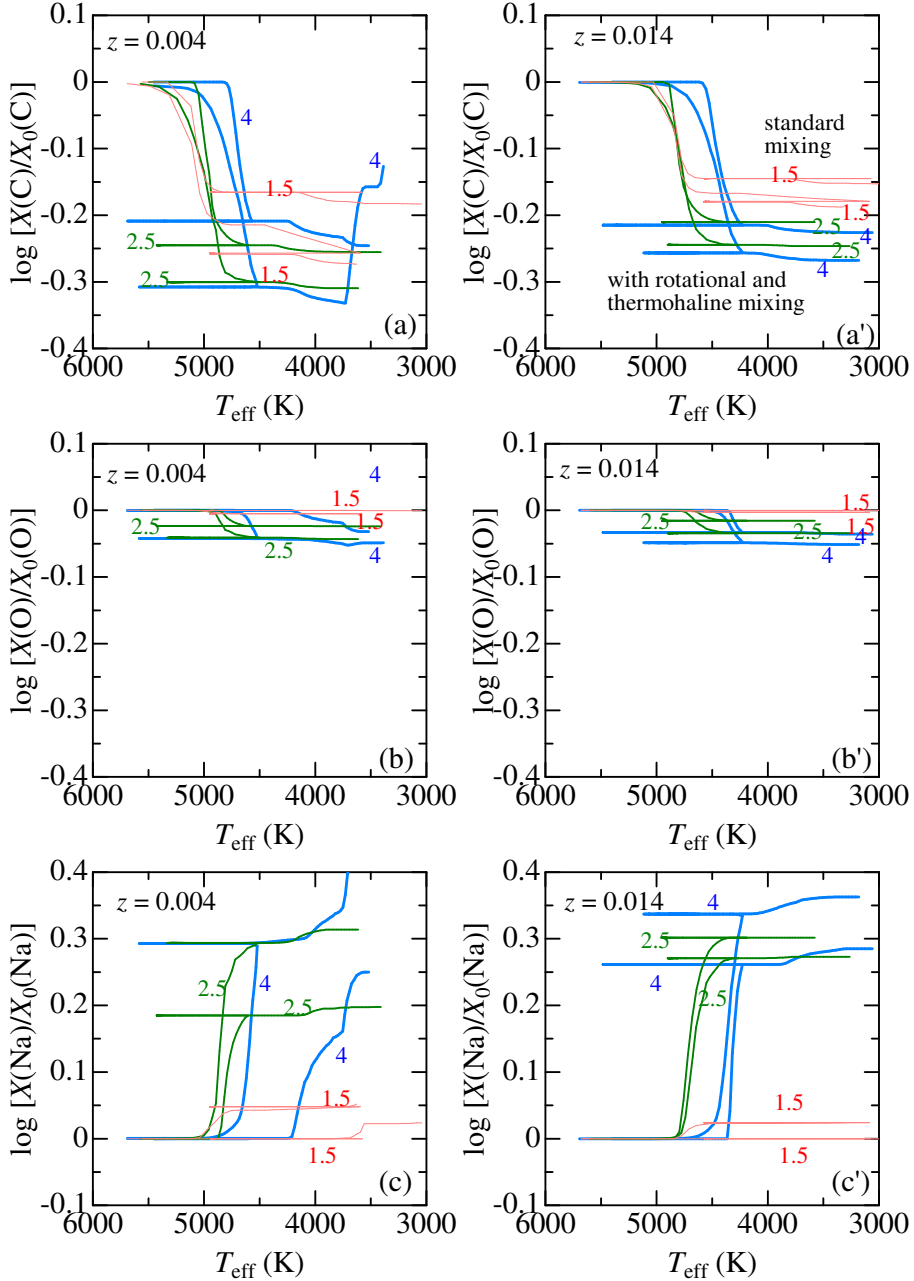}
  \end{center}
\caption{
Lagarde et al.'s (2012) theoretically simulated $\log [X/X_{0}]$
(logarithmic mass fraction ratio of the relevant element at the surface 
relative to the initial value) plotted against $T_{\rm eff}$, 
where top, middle, and bottom panels are for 
C (= $^{12}$C+$^{13}$C+$^{14}$C $\simeq ^{12}$C), O (= 
$^{16}$O+$^{17}$O+$^{18}$O $\simeq ^{16}$O), and Na (=$^{23}$Na), respectively. 
The left panels are for $z = 0.04$ ($0.3 \times$ solar metallicity)
and the right are for  $z = 0.14$ ($1 \times$ solar metallicity).
The results corresponding to three stellar masses of 1.5, 2.5, 
and 4.0~$M_{\odot}$ are shown here, which are discriminated 
by line thickness (thin orange line, normal green line, and 
thick blue lines, respectively).
Here, we restricted the data only to those of the well-evolved red-giant stage 
satisfying the conditions of $T_{\rm eff} < 5700$~K and $age > 10^{7.5}$~yr.
Note that two kinds of curves are shown corresponding to different treatments 
of envelope mixing; i.e., standard treatment and treatment including rotational 
and thermohaline mixing. Although these two sets are drawn in the same line-type,
they are discernible as the latter generally shows larger anomaly 
(and the appearance of peculiarity tends to begin earlier; i.e., at 
higher $T_{\rm eff}$) as compared to the former. 
}
\end{figure}

\end{document}